\documentclass[sigconf]{acmart}
\usepackage{caption}
\usepackage{subfigure}
\usepackage[T1]{fontenc}

\usepackage{amsmath,amsfonts}
\usepackage{algorithmic}
\usepackage{graphicx}
\usepackage{textcomp}
\usepackage{xcolor}
\usepackage{color, colortbl}
\definecolor{Gray}{gray}{0.9}
\usepackage{amsmath}
\usepackage{pifont}

\usepackage{makecell}
\usepackage{wrapfig}
\usepackage{picinpar}

\usepackage[utf8]{inputenc} %
\usepackage{hyperref}       %
\usepackage{url}            %
\usepackage{booktabs}       %
\usepackage{amsfonts}       %
\usepackage{nicefrac}       %
\usepackage{microtype}      %

\usepackage{algorithmic}
\usepackage{graphicx}
\usepackage{textcomp}
\usepackage{xcolor}
\usepackage{multirow}
\usepackage{algorithm}
\usepackage{bm}
\usepackage{subfigure}
\usepackage{caption}

\usepackage{CJKutf8}
\usepackage[T1]{fontenc}
\AtBeginDocument{%
  \providecommand\BibTeX{{%
    \normalfont B\kern-0.5em{\scshape i\kern-0.25em b}\kern-0.8em\TeX}}}

\copyrightyear{2024}
\acmYear{2024}
\setcopyright{acmlicensed}\acmConference[SIGIR '24]{Proceedings of the 47th International ACM SIGIR Conference on Research and Development in Information Retrieval}{July 14--18, 2024}{Washington, DC, USA}
\acmBooktitle{Proceedings of the 47th International ACM SIGIR Conference on Research and Development in Information Retrieval (SIGIR '24), July 14--18, 2024, Washington, DC, USA}
\acmDOI{10.1145/3626772.3657693}
\acmISBN{979-8-4007-0431-4/24/07}

\begin{document}

\title[CaseLink]{CaseLink: Inductive Graph Learning for Legal Case Retrieval}

\author{Yanran Tang}
\email{yanran.tang@uq.edu.au}
\affiliation{\institution{The University of Queensland}
  \city{Brisbane}
  \country{Australia}}
\author{Ruihong Qiu}
\email{r.qiu@uq.edu.au}
\affiliation{\institution{The University of Queensland}
  \city{Brisbane}
  \country{Australia}}
\author{Hongzhi Yin}
\email{h.yin1@uq.edu.au}
\affiliation{\institution{The University of Queensland}
  \city{Brisbane}
  \country{Australia}}
\author{Xue Li}
\email{xueli@eecs.uq.edu.au}
\affiliation{\institution{The University of Queensland}
  \city{Brisbane}
  \country{Australia}}
\author{Zi Huang}
\email{helen.huang@uq.edu.au}

\affiliation{\institution{The University of Queensland}
  \city{Brisbane}
  \country{Australia}}

\renewcommand{\shortauthors}{Yanran Tang, Ruihong Qiu, Hongzhi Yin, Xue Li, \& Zi Huang}

\begin{abstract}
In case law, the precedents are the relevant cases that are used to support the decisions made by the judges and the opinions of lawyers towards a given case. This relevance is referred to as the case-to-case reference relation. To efficiently find relevant cases from a large case pool, retrieval tools are widely used by legal practitioners. Existing legal case retrieval models mainly work by comparing the text representations of individual cases. Although they obtain a decent retrieval accuracy, the \textbf{\textit{intrinsic case connectivity}} relationships among cases have not been well exploited for case encoding, therefore limiting the further improvement of retrieval performance. In a case pool, there are three types of case connectivity relationships: the case reference relationship, the case semantic relationship, and the case legal charge relationship. Due to the inductive manner in the task of legal case retrieval, using case reference as input is not applicable for testing. Thus, in this paper, a CaseLink model based on inductive graph learning is proposed to utilise the intrinsic case connectivity for legal case retrieval, a novel Global Case Graph is incorporated to represent both the case semantic relationship and the case legal charge relationship. A novel contrastive objective with a regularisation on the degree of case nodes is proposed to leverage the information carried by the case reference relationship to optimise the model. Extensive experiments have been conducted on two benchmark datasets, which demonstrate the state-of-the-art performance of CaseLink. The code has been released on ~\url{https://github.com/yanran-tang/CaseLink}.
\end{abstract}



\maketitle

\section{Introduction}
In case law, a legal system widely applied by countries like Australia, the United Kingdom, the United States etc., the judicial reasons of a judgement for a given case are critically based on precedents, which are the legal cases relevant to the given case~\cite{precedent}. Legal case retrieval (LCR) model is an efficient tool in helping legal practitioners to effectively search for relevant cases from large databases.
\begin{figure}[!t]
\centering
\includegraphics[width=\linewidth]{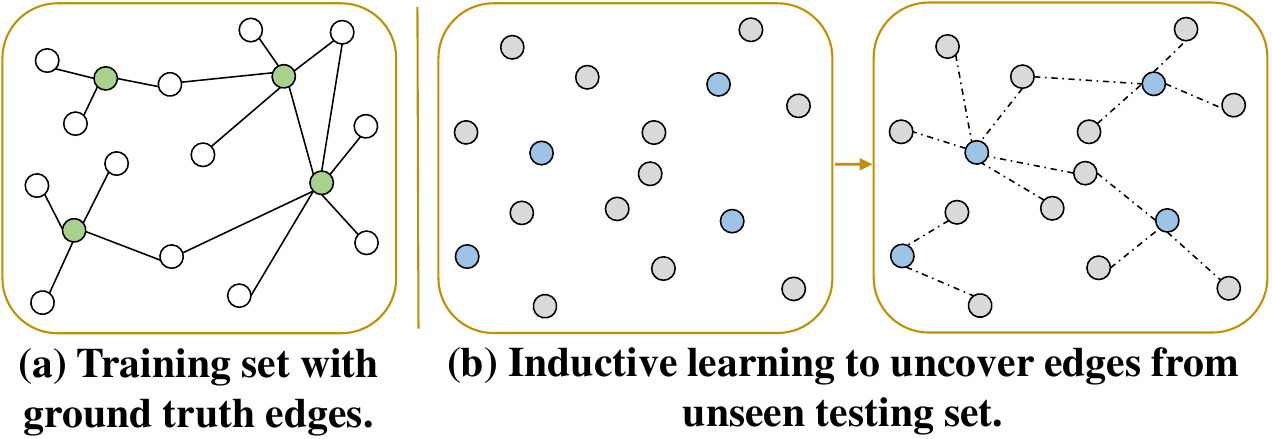}
\caption{The inductive nature of case reference in legal case retrieval. (a) During training, a labelled dataset contains query cases (green nodes), candidate cases (white nodes), and the ground truth reference between queries and candidates (solid edges). For simplicity, edges are denoted as undirected. (b) During inductive testing, given an unlabelled and unseen dataset with new query cases (blue nodes) and candidate cases (grey nodes), legal case retrieval models are expected to uncover case references in dashed edges.}
\label{fig:inductive}
\end{figure}

Generally, existing LCR models can be divided into two types: statistical models and neural network models. Statistical models, such as BM25~\cite{BM25}, TF-IDF~\cite{TF-IDF} and LMIR~\cite{LMIR}, use term frequency in documents and corpus to measure the relevance between cases. For neural network models~\cite{Law2Vec,Lawformer,MTFT-BERT,MVCL,LEGAL-BERT,JOTR,DoSSIER,RPRS,NOWJ,UA@COLIEE2022,IOT-Match,Law-Match,LEDsummary,BM25injtct,LeiBi,LEVEN,JNLP@COLIEE2019,CL4LJP,QAjudge,query_conversational_agent, ConversationalAgent,BERT-PLI,SAILER,promptcase,casegnn,gear,caseencoder,cfgl}, the similarity between cases are obtained based on  individually encoded representations of case texts. Typically, the encoding involves language models (BERT-PLI~\cite{BERT-PLI}, SAILER~\cite{SAILER} and PromptCase~\cite{promptcase}), or graph neural networks~\cite{GCN} (CaseGNN~\cite{casegnn}), to leverage the semantics in the case text.

In the LCR scenario, the ultimate goal is to uncover the connection relationship among cases as in Figure~\ref{fig:inductive}, where the connection relationship has not been thoroughly investigated by existing methods. In this work, it is argued that there are three types of \textbf{\textit{intrinsic case connectivity}} relationships among legal cases: (1) the case reference relationship, originating from precedent cases; (2) the case semantic relationship, calculated based on the semantic similarity between cases; and (3) the case legal charge relationship, standing for the relationship between a case and the charges of the legal case system, as well as the higher order relationship between cases of same or similar legal charges. These connectivity relationships are essential for LCR since the relevance can be uncovered from them. Existing LCR neural network models mainly focus on individual case text representation without sufficiently utilising these intrinsic case connectivity relationships within the case pool. This individual case encoding paradigm is not effective enough to support the legal case retrieval, where the case relationships can be complicated with various connectivity. Moreover, these relationships are not easy to discover and utilise in a straightforward style for legal case retrieval. It is challenging yet meaningful to effectively utilise the intrinsic case connectivity in LCR models.

In light of the above discussions, the main idea is to convert a pool of stand-alone cases into a structured graph, which will be further transformed into a graph with the expected case reference structure. Due to the inductive manner in the task of LCR, the case reference relationship is the ground truth label in LCR, which means that case reference is not applicable to be the input for models. Therefore, in this paper, a novel CaseLink framework is proposed to effectively perform LCR in an inductive graph learning manner. Firstly, a Global Case Graph (GCG) is constructed by leveraging the pairwise connectivity relationships of cases with case-to-case and case-to-charge relationship. Specifically, the term-frequency level and the semantic level similarities between cases are calculated to provide the potential case-to-case relationship while the legal charges of each case is extracted for case-to-charge relationship. The extracted case-to-case and case-to-charge relationship jointly offer the pairwise connectivity of cases. To leverage the connectivity relationships in GCG, a graph neural network module is developed to perform graph representation learning. Finally, a new objective function consisting of contrastive loss and degree regularisation is designed to train the CaseLink. Two benchmark datasets COLIEE2022~\cite{COLIEE2022} and COLIEE2023~\cite{COLIEE2023} are experimented in this paper. The empirical results demonstrate the state-of-the-art performance of CaseLink and the effectiveness on LCR task. The main contributions of this paper are summarised as follows:
\begin{itemize}
\item This paper investigates the case reference relationship in the LCR problem and transforms the traditional LCR paradigm into an inductive graph learning procedure.
\item A CaseLink framework is proposed to exploit the case reference information. It consists of a graph neural network-based pipeline to learn the latent connectivity among cases with a Global Case Graph and a degree regularisation.
\item Extensive experiments conducted on two benchmark datasets demonstrate the state-of-the-art performance of CaseLink and the effectiveness of the global graph-based design.
\end{itemize}

\section{Related Work}

\subsection{Legal Case Retrieval}
Legal case retrieval aims to retrieve relevant cases from a large database given a query legal case, belonging to query-by-document category. Recent LCR methods can be mainly divided into two types: statistical models~\cite{TF-IDF,BM25,LMIR} and neural network models~\cite {Law2Vec,Lawformer,MTFT-BERT,MVCL,LEGAL-BERT,JOTR,DoSSIER,RPRS,NOWJ,UA@COLIEE2022,IOT-Match,Law-Match,LEDsummary,BM25injtct,LeiBi,LEVEN,JNLP@COLIEE2019,CL4LJP,QAjudge,query_conversational_agent, ConversationalAgent,BERT-PLI,SAILER,promptcase,casegnn,caseencoder,gear,queryreformaulation}. For statistical models, most methods rely on using term frequency to measure the similarity between cases, such as TF-IDF~\cite{TF-IDF}, BM25~\cite{BM25} and LMIR~\cite{LMIR}. These methods are convenient for conducting calculations since there is only term frequency counting without optimisation and inference of large neural network models. One of the drawbacks of these statistical models is that only using term frequency cannot effectively represent the relevance between cases given the complexity of the individual cases. For the neural network models, most of the existing work relies on encoding the case text into a high-dimensional vector with neural networks of various structures. For example, BERT~\cite{BERT} is a typical encoder that generates a representation for the case text. Given that there is an input length limit (e.g., 512 tokens) for BERT while a case can generally contain thousands of words, most BERT-based LCR methods aim to bypass this restriction while maintaining the quality of the case representation. BERT-PLI~\cite{BERT-PLI} proposes a paragraph-level interaction module by dividing the case into multiple paragraphs and measuring the similarity between cases by aggregating the paragraph-level similarity. SAILER~\cite{SAILER} chooses to truncate the overlong case text. One recent model, Gear~\cite{gear}, makes use of generative retrieval with legal judge prediction, which is similar to the case reasoning generation pre-training in SAILER. IOT-Match~\cite{IOT-Match} makes use of sentence embeddings and calculates the similarity with an optimal transport distance. PromptCase~\cite{promptcase} relies on generating a summary of the legal fact and identifying the legal issue sections together with a prompt reformulation to obtain an effective input representation. CaseGNN~\cite{casegnn} further transforms the unstructured case text into a structured case graph for each case and encodes the case with a dedicated graph neural work model. In general, most of these existing LCR models put their effort into using the case text to encode the individual cases. In contrast, the proposed CaseLink targets at the LCR problem from the perspective of exploiting the case reference information with latent connectivity of cases.

\subsection{Graph Neural Networks}
\label{sec:gnn-related}
Graph neural networks (GNNs) are an effective tool for representation learning on graph data~\cite{GCN,GAT,GraphSAGE}. Graph convolutional network (GCN) develops a convolution operation on graph data in a transductive setting~\cite{GCN}. Graph attention network (GAT) extends the GCN layer with an attention module~\cite{GAT}. GraphSAGE further designs an inductive model with neighbour sampling~\cite{GraphSAGE}. Graph has been used for many applications~\cite{hypergraph,inductivegraphcondensation,imgagn,diseaseprediction,crowdsourcing,fgnn,fgnnj,gag,posrec,cat,puma}.

\paragraph{GNN has been introduced into topics related to legal case understanding in recent years~\cite{LegalGNN,SLR,casegnn}.} CaseGNN~\cite{casegnn} firstly uses graph structure to represent an individual case and develops a weighted graph attention layer to encode the case semantics for the LCR problem. SLR~\cite{SLR} and CFGL-LCR~\cite{cfgl} apply graph representation learning on the external legal knowledge graph in addition to the general language model for LCR. For legal case recommendation, a recommendation scenario with users involved, LegalGNN~\cite{LegalGNN} develops a bipartite user-case graph recommendation method. Different from these graph-based methods in the legal domain, CaseLink focuses on the learning of latent relationships among legal cases rather than putting effort into individual case learning or user modelling.

\paragraph{GNN has also been an effective tool in other information retrieval or language-related topics.} In GNN-DocRetrieval~\cite{gnn-doc}, candidate documents are connected to a word graph to obtain document representations for COVID question-answering retrieval. AggRanker~\cite{aggranker} used the ground truth query-document label from user history to construct a behaviour graph for an online search engine. While in text classification, TextGCN~\cite{textgcn}, HeteGCN~\cite{hetegcn} and InducT-GCN~\cite{inductgcn} all develop the graph construction relying on document-word relationship to obtain text representations. Compared with these graph-based methods in other domains, CaseLink designs a novel document-document graph without relying on the ground truth query-document relationship nor the document-word relationship, which is more effective in an inductive setting, where test queries and test documents are not available during training. From a practical perspective, legal cases generally contain thousands of words, making connecting documents to words to form a graph ineffective and impractical due to the heavy and dense connections.

\section{Preliminary}
In the following, a lowercase letter represents a scalar or a sequence of words, a bold lowercase letter represents a vector, a bold uppercase letter represents a matrix, and a scripted uppercase letter represents a set.

\subsection{Task Definition}
\label{sec:task}
With a set of $n$ cases, $\mathcal{D}=\{d_1,d_2,...,d_n\}$, for a query case $q\in\mathcal{D}$, the task of legal case retrieval is to retrieve a set of relevant cases $\mathcal{D}^* = \{d^*_i| d^*_i \in \mathcal{D} \wedge relevant (d^*_i, q) \}$ from $\mathcal{D}$, where $relevant (d^*_i, q)$ represents that $d^*_i$ is a relevant case of query case $q$. Specifically, the relevant cases denote the precedents in legal domain, which are previous cases that was referred by the query case.

Note that in widely used benchmarks for LCR, such as COLIEE2022~\cite{COLIEE2022} and COLIEE2023~\cite{COLIEE2023}, the training data and the testing data do not overlap for queries and candidates, which makes the LCR task as an inductive retrieval as shown in Figure~\ref{fig:inductive}.

In this paper, a legal case will be formulated as a node and a pool of legal cases will be transformed into a graph. Therefore, the term ``case'' and the term ``node'' will be used interchangeably below.

\subsection{Transductive and Inductive Learning}
\label{sec:transVSinduct}
Note that the definitions of transductive and inductive learning in graph representation learning are different from the transductive and inductive learning in retrieval. (1) Given a graph, the transductive learning indicates that the testing nodes are available without label during training. While for the inductive learning, the testing nodes are unavailable during training, and will be connected to the training graph during testing~\cite{GraphSAGE}. One exception for the inductive learning in graph is the Graph-less Neural Network~\cite{glnn}, where the testing nodes will not be connected to any existing graph and the authors develop linear models without using the structure for testing. (2) For information retrieval, the transductive learning allows the training and the testing to share the same candidate pool. The inductive learning will not have any data overlap between training and testing. In LCR, this implies that there is no available structural information, i.e., case reference information, among cases in testing.

\section{Method}

\begin{figure}[!t]
\centering
\includegraphics[width=0.7\linewidth]{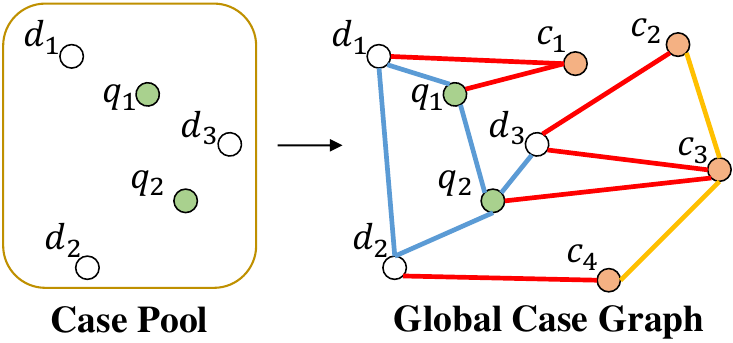}
\caption{An illustration of Global Case Graph. Green nodes are query cases $q_1$ and $q_2$, white nodes are candidate cases $d_1\sim d_3$ and orange nodes are legal charges $c_1\sim c_4$. Solid lines are edges, including case-case edges in blue, case-charge edges in red and charge-charge edges in yellow.}
\label{fig:gcg}
\end{figure}

\begin{figure*}[!t]
\centering
\includegraphics[width=0.7\linewidth]{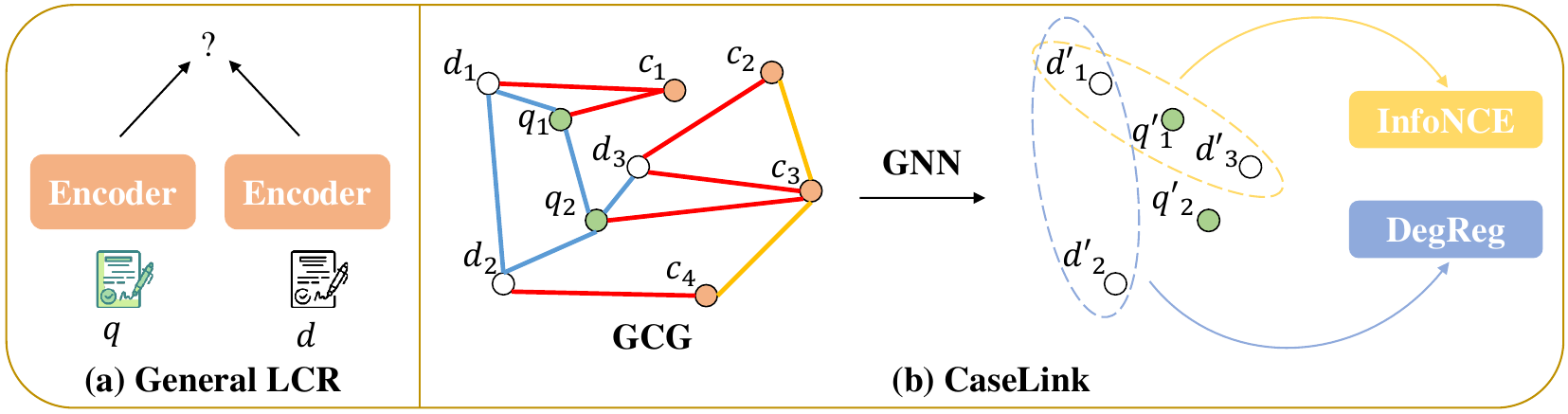}
\caption{The comparison between typical LCR models and CaseLink. (a) Existing LCR models generally apply a text encoder to query and candidate individually. The LCR prediction is obtained by perform nearest neighbour search on these non-interactive encodings. (b) The overall framework of CaseLink. During training, the training queries, the training candidates and the charges are transformed into a Global Case Graph (GCG). A graph neural network (GNN) module will conduct the node feature update for the GCG. The updated query and candidate node features will be fed into the contrastive learning (InfoNCE) objective and the degree regularisation (DegReg) objective to train the CaseLink model. During inference, the testing queries, the testing candidates and the charges are transformed into another GCG. After obtaining the updated node features with the GNN module, the retrieval result is achieved by the nearest neighbour search based on the similarity among these case node features.}
\label{fig:caselink}
\end{figure*}

\subsection{Global Case Graph}
\label{sec:GCG}
Global Case Graph (GCG) aims to thoroughly utilised the case reference information to construct an informative case graph. To construct a GCG, the edges between case nodes are selected by using traditional statistical retrieval method and the feature of case nodes are generated by using a legal case encoder. The illustration of an exemplar GCG is demonstrated in Figure~\ref{fig:gcg}.

In this paper, a GCG is denoted as $G = (\mathcal{V}, \mathcal{E})$, where $\mathcal{V}$ and $\mathcal{E}$ is the set of nodes and edges respectively in the GCG. Specifically, there are two node types in GCG, the case node $d\in \mathcal{V}$ with the the node feature $\textbf{x}_d \in \mathbb{R}^d$, and the charge node $c\in \mathcal{V}$ with the feature $\textbf{x}_c \in \mathbb{R}^d$. The case node refers to the case in the case pool $\mathcal{D}$ while the charge node comes from the legal charge extracted from these 
cases. Although there are two types of nodes, the graph learning scenario is still framed into a homogeneous situation since these node features both originate from text encodings. If there is an edge between any two nodes $u$ and $v$, the edge is denoted as $e_{uv}\in \mathcal{E}$. The definition of edges is provided below.

\subsubsection{Case Node}
\label{sec:casenode}
In GCG, $n$ cases in the case pool $\mathcal{D}$, including both query and candidate cases, will be converted into $n$ case nodes as shown in the green (query node) and the white (candidate node) circles of Figure~\ref{fig:gcg}. The feature of every node is from an encoder as:
\begin{equation}
\label{eq:casenode}
    \mathbf{x}_{d} = \text{Encoder}_{\text{case}}(t_{d}),
\end{equation}
where $t_{d}$ is the text of case $d$ and $\textbf{x}_{d} \in \mathbb{R}^d$ is the node feature of case node $d$ obtained from the encoder. $\text{Encoder}_{\text{case}}$ can be any encoder that can encode the case text into a case representation, such as a language model like BERT~\cite{BERT}, SAILER~\cite{SAILER} or PromptCase~\cite{promptcase}, and a graph neural network encoder like CaseGNN~\cite{casegnn}.

\subsubsection{Charge Node}
\label{sec:chargenode}
Considering the latent connectivity within the case pool, legal charges are important sources to link cases together. Since the cases in the experimented benchmarks all come from the Federal Court of Canada case laws, namely COLIEE2022~\cite{COLIEE2022} and COLIEE2023~\cite{COLIEE2023}, the charges used in the experiments can be found in a list of the Federal Courts Act and Rules of Canada\footnote{\url{https://www.fct-cf.gc.ca/en/pages/law-and-practice/acts-and-rules/federal-court/}}, denoted as $\mathcal{C}=\{c_1,c_2,...,c_m\}$. The features of charge nodes are encoded by an encoder with the text of charges, which is defined as:
\begin{equation}
\label{eq:casenode}
    \mathbf{x}_{c} = \text{Encoder}_{\text{charge}}(t_{c}),
\end{equation}
where $t_{c}$ is the text of charge $c$, $\textbf{x}_{c} \in \mathbb{R}^d$ is the node feature of charge node $c$ obtained from the encoder. Similar to case nodes, $\text{Encoder}_{\text{charge}}$ can be any encoder that encodes the short charge phrases into a charge representation, such as a language model like BERT~\cite{BERT}, SAILER~\cite{SAILER} or PromptCase~\cite{promptcase}.

\subsubsection{Case-Case Edge}
\label{sec:caseedge}
According to the case relationship within the case pool, it is desired to link the cases that have intrinsic connectivity to be neighbouring nodes for a more effective message passing of case nodes. Thus, the pairwise BM25~\cite{BM25} score between every cases is calculated. Due to the large number of cases, only top $k$ highest BM25 similarity scores cases are selected and linked as blue edges in Figure~\ref{fig:gcg}. Although BM25 is a non-symmetric measurement, which essentially makes the case-case edge directed, for the simplicity of CaseLink framework, the case-case edge is converted into symmetric. The final adjacency matrix of case-case edge $\textbf{A}_{d} \in \mathbb{R}^{n\times n}$ ($n$ cases in case pool in total) is denoted as:
\begin{equation}
\label{eq:d-edge}
\textbf{A}_{d_{ij}}=\left\{
\begin{aligned}
1 \quad & \text{for} & \text{TopK}(\text{BM25}(t_{d_{i}},t_{d_{j}}|d_{i},d_{j} \in \mathcal{D})), \\
0 \quad & \text{for} & \text{Others}, 
\end{aligned}
\right.
\end{equation}
where $d_{i}$ and $d_{j}$ are two cases in the case pool $\mathcal{D}$. BM25~\cite{BM25} is a statistical model that calculates the text similarity using term-frequency. Given a list of cases BM25 score, TopK is a function that returns a set $\mathcal{K}$ of top $k$ cases in the case pool $\mathcal{D}$, denoted as $\mathcal{K}=\{d_1,d_2,...,d_k|d_k \in \mathcal{D}\}$. Either $\text{BM25}(t_{d_{i}},t_{d_{j}})$ or $\text{BM25}(t_{d_{j}},t_{d_{i}})$ in the top $K$ will result in edges $\textbf{A}_{d_{ij}}$ and $\textbf{A}_{d_{ji}}$, which is equivalent to a logical OR operation and makes the adjacency matrix $\textbf{A}_d$ symmetric.

\subsubsection{Charge-Charge Edge}
\label{sec:chargeedge}
Within a legal system, there are natural relationships between different charges by considering that multiple charges may simultaneously appear in the same case. When two charges are linked by an edge, they may have a higher probabilities to both appear in two similar cases than other separated charges. The symmetric adjacency matrix of charge-charge edge $\textbf{A}_{c}\in\mathbb{R}^{m\times m}$ ($m$ is number of charges) defined as:
\begin{equation}
\label{eq:c-edge}
\textbf{A}_{c_{ij}}=\left\{
\begin{aligned}
1 \quad & \text{for} & \text{Sim}(\textbf{x}_{c_{i}},\textbf{x}_{c_{j}}|c_{i},c_{j} \in \mathcal{V})>\delta, \\
0 \quad & \text{for} & \text{Others}, 
\end{aligned}
\right.
\end{equation}
where $c_{i}, c_{j}$ are two charge nodes in $\mathcal{V}$ with the node features $\textbf{x}_{c_{i}} \in \mathbb{R}^d$, $\textbf{x}_{c_{j}} \in \mathbb{R}^d$ from Section~\ref{sec:chargenode}. The Sim function is used to calculated the similarity between two charge nodes, which can be dot product and cosine similarity. $\delta$ is the threshold value for controlling the number and equality of charge-charge edge.

\subsubsection{Case-Charge Edge}
Clarifying the legal charge of a case is essential for legal practitioners to understand the case itself and identify relevant cases. The case-charge edge exists when a charge appears in a given case, which means that there is a natural one-way inclusion relationship between case and charge. The adjacency matrix of case-charge edge $\textbf{A}_b\in\mathbb{R}^{m\times n}$ is denoted as:
\begin{equation}
\label{eq:cd-edge}
\textbf{A}_{b_{ij}}=\left\{
\begin{aligned}
1 \quad & \text{for} & t_{c_{i}} \ \text{appears in} \ t_{d_{j}}, \\
0 \quad & \text{for} & \text{Others},
\end{aligned}
\right.
\end{equation}
where $t_{c_{i}}$ is the text of charge $i$ and $t_{d_{j}}$ is the text of case $j$.

\subsubsection{Overall Adjacency Matrix}
The edges of GCG including case-case, charge-charge and case-charge edges are undirected and unweighted. The overall adjacency matrix $\textbf{A} \in \mathbb{R}^{(n+m)\times (n+m)}$ is:
\begin{equation}
 \textbf{A} = \begin{bmatrix}
 \textbf{A}_{d}  & \textbf{A}_{b}^T \\
 \textbf{A}_{b} & \textbf{A}_{c} 
 \end{bmatrix},
\end{equation}
where $\textbf{A}_{b}^T$ is the transpose matrix of adjacency matrix $\textbf{A}_{b}$. The overall adjacency matrix, $\textbf{A}$, is symmetric.
 
\subsection{Graph Neural Network Module}
\label{sec:gnn}
With the constructed GCG, a GNN module is leveraged to aggregate the information in nodes and edges that contain the intrinsic case connectivity to generate a comprehensive case representation.

\subsubsection{Graph Neural Network}
\label{sec:GNN}
After $k-1$ layers of GNN calculations, the output feature of node $v$ in the $k$-th layer is denoted as:
\begin{equation}
    \textbf{h}_{v}^{k}=\text{GNN}(\textbf{h}_v^{k-1}, \textbf{h}_u^{k-1}: u \in \mathcal{N}(v)),
\end{equation}
where $\textbf{h}_v^{k-1} \in \mathbb{R}^d$ and $\textbf{h}_u^{k-1} \in \mathbb{R}^d$ are the node representations of $v$ and $u$ at $(k-1)$-th layer and $\mathcal{N}(v)$ is the neighbour node set of node $v$. $\textbf{h}_v^{k} \in \mathbb{R}^d$ is the output of $k$ GNN layers and also the feature of node $v$ in $k$-th layer. The initialisation of $\textbf{h}^{0}$ is assigned with the encodings $\textbf{x}_d$ and $\textbf{x}_c$ from Section~\ref{sec:casenode} and~\ref{sec:chargenode}. GNN can be any graph neural network model that can exploit the graph structure information to generate a representative node embeddings, such as GCN~\cite{GCN}, GAT~\cite{GAT} or GraphSAGE~\cite{GraphSAGE}.

\subsubsection{Residual Connection}
Moreover, to effectively utilise the input information of cases and charges, a residual connection is used in the end after GNN layer to obtain final node features $\textbf{h}_v$:
\begin{equation} 
    \textbf{h}_v=\textbf{h}_{v}^{k}+\mathbf{x}_{v},
\end{equation}

\subsection{Objective Function}
\label{sec:obj}

\subsubsection{Contrastive Learning of Query}
The task of LCR is to distinguish the relevant cases from a huge collection of cases for a given query. This is also the same goal as the contrastive learning to pull the positive samples closer while push the negative samples far away used in retrieval task~\cite{CL4LJP,casegnn,SAILER,MVCL}. Therefore, in this paper, the main objective of query is designed in an InfoNCE~\cite{infonce} style as:
\begin{equation}
\label{eq:infonce}
\begin{split}
&\ell_{\text{InfoNCE}}=\\
&-\text{log}\frac{e^{\frac{(s(\mathbf{h}_q,\mathbf{h}_{d^+}))}{\tau}}}{e^{\frac{(s(\mathbf{h}_q,\mathbf{h}_{d^+}))}{\tau}}+\sum\limits^{n_e}_{i=1}e^{\frac{(s(\mathbf{h}_q,\mathbf{h}_{d^{easy-}_i}))}{\tau}}+\sum\limits^{n_h}_{i=1}e^{\frac{(s(\mathbf{h}_q,\mathbf{h}_{d^{hard-}_i}))}{\tau}}},
\end{split}
\end{equation}
where $q$ is the query case and $\mathcal{D}$ is the case pool that includes both relevant cases $d^+$ and irrelevant cases $d^-$. $s$ is a similarity metric that can compare the similarity between two vectors, such as dot product or cosine similarity. $n_e$ and $n_h$ are the number of easy negative sample $d^{easy-}$ and hard negative sample $d^{hard-}$, respectively. $\tau$ is the temperature coefficient. In the training processing, the positive samples are the ground truth label of training dataset, the easy negative samples are randomly sampled form the case pool $\mathcal{D}$ as well as simultaneously using the in-batch samples of other queries. Specially, to effectively guide the training with making use of harder samples, the hard negative samples are selected based on the BM25~\cite{BM25} relevance score. If a case have a high BM25 relevance score to the query while is still not a positive case, such a case is sampled as a hard negative case. 

\subsubsection{Degree Regularisation of Candidate}
\label{sce:regloss}
When only using contrastive objective, there is limited training signal for training candidates in the case pool. Thus, a degree regularisation is proposed to minimise the case node degree for candidate nodes, which can serve as the training signal for candidates as well as the regularisation to meet the real-world requirement that these cases should be just related to a small amount of cases in the pool and the case reference is sparsely connected. To derive the degree regularisation of candidates, the pseudo adjacency matrix of the nodes with updated features after GNN calculation is defined as:
\begin{equation}
    \hat{\textbf{A}}_{ij} = \text{cos}(\mathbf{h}_{i},\mathbf{h}_{j}),
\end{equation}
where $\mathbf{h}_{i}$ and $\mathbf{h}_{j}$ are the updated features of case node $i$ and $j$ in the case pool $\mathcal{D}$. The matrix $\hat{\textbf{A}}\in\mathbb{R}^{n\times n}$ indicates a fully connected situation. And the degree regularisation is conducted on this pseudo adjacency matrix $\hat{\textbf{A}}$ only for candidate cases:
\begin{equation}
    \ell_{\text{DegReg}} = \sum\limits^o_{i=1}\sum\limits^n_{j=1}(\hat{\textbf{A}}_{ij}),         
\end{equation}
where $o$ is the number of candidate cases in $\mathcal{D}$. 

\subsubsection{Overall Objective}
\label{sce:loss}
During training, the overall objective is:
\begin{equation}
\label{eq:loss-overall}
    \ell = \ell_{\text{InfoNCE}}+\lambda\cdot\ell_{\text{DegReg}},
\end{equation}
where $\lambda$ is the coefficient for the scale of the degree regularisation.

\subsection{Inference}
\label{sec:inf}
Given a testing case pool $\mathcal{D}_{\text{text}}$, the relevance score $s_{(q,d)}$ between testing query case $q$ and the candidate case $d$ is calculated as:
\begin{equation} 
\label{eq:inference}
    s_{(q,d)} = \text{cos} (\mathbf{h}_{q}, \mathbf{h}_{d}),   
\end{equation}
where $\mathbf{h}_{q}$ and $\mathbf{h}_{d}$ are the representations of query $q$ and candidate $d$ from CaseLink. Candidates with top ranking scores are retrieved.

\section{Discussion}

\subsection{Inductive Learning}
As described in Section~\ref{sec:task} and~\ref{sec:transVSinduct}, the LCR problem is under the inductive learning nature where testing queries and candidates are both unavailable during training. This puts various restrictions on the design of using the reference connection in the graph structure.

\subsubsection{Node Initialisation} Considering related research areas such as recommender systems, which also use the graph structure to connect the data points, an effective initialisation for node features is using the ID embedding~\cite{ngcf}. This is under the assumption that the coverage of users and items during training and testing will not change, which falls exactly into the transductive learning. However, in LCR problem, the testing queries and candidates are all unavailable during training, which prevents from using the case ID as node features. Otherwise, it will become a cold-start situation.

\subsubsection{Edge Design} A straightforward solution to build an input graph to utilise the reference relationship of cases for training an LCR model is to link the cases that has the real reference relationship. However, in LCR problem, the inductive learning nature indicates that during testing, there is no available reference relationship to build such a testing graph. Therefore, it is impossible to have a same data distribution for both training and testing if the real reference relationships are used as edges.

\subsection{Graph Extensions}
In the current design of GCG, the edges are converted into undirected and unweighted for simplicity. There are different chances to extend these edge designs to include direction and weight.

\subsubsection{Directed Graph.} The current case-case edge comes from the BM25 score between cases. Note that BM25 score is asymmetric, a natural extension of the case-case edge is to include the direction coming from BM25. For the case-charge edges, the inclusion relation between cases and charges can also be considered as directed. However, it is non-trivial to provide directions for charge-charge edges since there is no clear directed relationship between charges. A possible solution is to leverage external legal expert knowledge to define the relationship among charges.

\subsubsection{Weighted Graph.} For all the edges in the current design, they all come from a quantitative measurement, such as BM25 score, semantic similarity and inclusion relation. All of these measurements are meaningful in real number space, which gives the possibility to include the measured values as edge weights. The difficult of using these values as weights is that the values are not directly comparable. For example, the value of BM25 score can be greater than $1,000$ while the semantic similarity from a language model can be ranging between $-1$ and $1$. Even after normalisation, the straightforward consideration of these values as weights is not convincing. A possible solution is to transform the graph into a heterogeneous graph, which would apply different functions to the edges coming from different measurements.

\subsection{Relationship with Graph-based Text Classification Methods}
As described in the end of Section~\ref{sec:gnn-related}, there are a few graph-based methods working on the text classification topic~\cite{textgcn,hetegcn,inductgcn}. In this area, a short text in a dataset is being classified by the model. Recent methods consider each piece of text as a node in the graph and connect these text nodes mainly based on the case-to-word relationship. Although CaseLink also transforms the data point, legal case, into a node, there is no case-to-word relationship in the Global Case Graph because (1) a legal case generally has thousands of words compared with a few dozens of words in short texts from text classification; and (2) the simple text matching is not meaningful in LCR problem. Additionally, the case-to-case connection is important in CaseLink, while not appearing in these text classification methods.

\section{Experiments}
In this section, the experiment settings and results are described, which aims to answer the following research questions (RQs):
\begin{itemize}
    \item RQ1: How does CaseLink perform compared with the state-of-the-art LCR models?
    \item RQ2: How effective are different types of case connectivity information in CaseLink?
    \item RQ3: How does graph learning help with LCR in CaseLink?
    \item RQ4: How do hyper-parameter settings affect CaseLink?
\end{itemize}

\subsection{Setup}
\label{sec:setup}
\begin{table}[!t]\centering
\caption{Statistics of datasets.}\label{tab:dataset}
\small
    \begin{tabular}{c|cc|cc}
    \toprule
    \multirow{2}{*}{Datasets} &\multicolumn{2}{c|}{COLIEE2022} &\multicolumn{2}{c}{COLIEE2023} \\
    \cmidrule{2-5}
    &train &test &train &test \\\midrule
    \# Query &898 &300 &959 &319 \\
    \# Candidates &4415 &1563 &4400 &1335 \\
    \# Avg. relevant cases &4.68 &4.21 &4.68 &2.69 \\
    Avg. length (\# token) &6724 &6785 &6532 &5566 \\
    Largest length (\# token) &127934 &85136 &127934 &61965 \\
    \bottomrule
    \end{tabular}
    \vspace{-0.5cm}
\end{table}

\subsubsection{Datasets.}
To evaluate the CaseLink, two benchmark datasets, COLIEE2022~\cite{COLIEE2022} and COLIEE2023~\cite{COLIEE2023}, are used in the experiments. Both datasets come from the Competition on Legal Information Extraction/Entailment (COLIEE), where the cases are collected from the Federal Court of Canada. There are two main difference between them. On on hand, the cases in test sets and most of the training sets in two datasets are different. On the other hand, the average relevant cases numbers per query as shown in Table~\ref{tab:dataset} are different, resulting in different difficulties for finding ground truth label cases. For reasonable text encoding, the French appears in the cases of both datasets are removed. The evaluation is based on full candidate pool ranking instead of sample ranking, which makes the evaluation unbiased and more challenging. Although these two datasets focus on English LCR scenario, CaseLink can be easily extended to other languages with respective language encoders.

\subsubsection{Metrics.}
\label{metrics}
To evaluate the performance of the experiments, the metric of  precision (P), recall (R), Micro F1 (Mi-F1), Macro F1 (Ma-F1), Mean Reciprocal Rank (MRR), Mean Average Precision (MAP) and normalized discounted cumulative gain (NDCG) are selected as they are widely used in information retrieval task.Top 5 ranking results are evaluated based on based on the previous LCR works~\cite{promptcase,LeCaRD,SAILER}. All metrics are the higher the better.

\begin{table*}[!t]\centering
\caption{Overall performance on COLIEE2022 and COLIEE2023 (\%). Underlined numbers indicate the best baselines. Bold numbers indicate the best performance of all methods. Both one-stage and two-stage results are reported.}\label{tab:overall}
\resizebox{1\linewidth}{!}{
\begin{tabular}{l|ccccccc|ccccccc}
\toprule
\multirow{2}{*}{Methods} &\multicolumn{7}{c}{COLIEE2022} &\multicolumn{7}{c}{COLIEE2023}\\
\cmidrule{2-15}
&P@5 &R@5 &Mi-F1 &Ma-F1 &MRR@5 &MAP &NDCG@5 &P@5 &R@5 &Mi-F1 &Ma-F1 &MRR@5 &MAP &NDCG@5 \\\midrule
\midrule
\textbf{One-stage}\\
BM25 &17.9 &21.2 &19.4 &21.4 &23.6 &25.4 &33.6 &16.5 &30.6 &21.4 &22.2 &23.1 &20.4 &23.7\\
LEGAL-BERT &4.47 &5.30 &4.85 &5.38 &7.42 &7.47 &10.9 &4.64 &8.61 &6.03 &6.03 &11.4 &11.3 &13.6\\
MonoT5 &0.71 &0.65 &0.60 &0.79 &1.39 &1.41 &1.73 &0.38 &0.70 &0.49 &0.47 &1.17 &1.33 &0.61 \\
SAILER &16.6 &15.2 &14.0 &16.8 &17.2 &18.5 &25.1 &12.8 &23.7 &16.6 &17.0 &25.9 &25.3 &29.3\\
PromptCase &17.1 &20.3 &18.5 &20.5 &35.1 &33.9 &38.7 &16.0 &29.7 &20.8 &21.5 &32.7 &32.0 &36.2 \\
CaseGNN&\underline{35.5}$\pm$0.2 &\underline{42.1}$\pm$0.2 &\underline{38.4}$\pm$0.3 &\underline{42.4}$\pm$0.1 &\underline{66.8}$\pm$0.8 &\underline{64.4}$\pm$0.9 &\underline{69.3}$\pm$0.8 &\underline{17.7}$\pm$0.7 &\underline{32.8}$\pm$0.7&\underline{23.0}$\pm$0.5 &\underline{23.6}$\pm$0.5 &\underline{38.9}$\pm$1.1 &\underline{37.7}$\pm$0.8 &\underline{42.8}$\pm$0.7\\
\rowcolor{lightgray}
CaseLink (Ours)&\textbf{37.0}$\pm$0.1&\textbf{43.9}$\pm$0.1&\textbf{40.1}$\pm$0.1&\textbf{44.2}$\pm$0.1&\textbf{67.3}$\pm$0.5&\textbf{65.0}$\pm$0.2&\textbf{70.3}$\pm$0.1&\textbf{20.9}$\pm$0.3&\textbf{38.4}$\pm$0.6&\textbf{27.1}$\pm$0.3&\textbf{28.2}$\pm$0.3&\textbf{45.8}$\pm$0.5&\textbf{44.3}$\pm$0.7&\textbf{49.8}$\pm$0.4\\
\midrule
\midrule
\textbf{Two-stage}\\
SAILER &\underline{23.8} &25.7 &24.7 &25.2 &43.9 &42.7 &48.4 &19.6 &32.6 &24.5 &23.5 &37.3 &36.1 &40.8\\
PromptCase &23.5 &25.3 &24.4 &\underline{30.3} &41.2 &39.6 &45.1 &\underline{\textbf{21.8}} &36.3 &\underline{27.2} &26.5 &39.9 &38.7 &44.0\\
CaseGNN&22.9$\pm$0.1 &\underline{27.2}$\pm$0.1 &\underline{24.9}$\pm$0.1 &27.0$\pm$0.1 &\underline{54.9}$\pm$0.4 &\underline{54.0}$\pm$0.5 &\underline{57.3}$\pm$0.6 &20.2$\pm$0.2 &\underline{37.6}$\pm$0.5 &26.3$\pm$0.3 &\underline{27.3}$\pm$0.2 &\underline{45.8}$\pm$0.9 &\underline{44.4}$\pm$0.8 &\underline{49.6}$\pm$0.8\\
\rowcolor{lightgray}
CaseLink (Ours)&\textbf{24.7}$\pm$0.1&\textbf{29.1}$\pm$0.1&\textbf{26.8}$\pm$0.1&29.2$\pm$0.1&\textbf{56.0}$\pm$0.2&\textbf{55.0}$\pm$0.2&\textbf{58.6}$\pm$0.1&21.0$\pm$0.3&\textbf{38.9}$\pm$0.5&27.1$\pm$0.3&\textbf{28.2}$\pm$0.3&\textbf{48.8}$\pm$0.2&\textbf{47.2}$\pm$0.1&\textbf{52.6}$\pm$0.1\\
\bottomrule
\end{tabular}}
\end{table*}

\subsubsection{Baselines.}
\label{baselines}
Various state-of-the-art baselines are used in the experiments to evaluate the performance of CaseLink as follows:
\begin{itemize}
    \item \textbf{BM25}~\cite{BM25}: a traditional but strong retrieval benchmark that utilises the term frequency to measure text similarity.
    \item \textbf{LEGAL-BERT (2020)}~\cite{LEGAL-BERT}: a language model that is pre-trained on large legal corpus.
    \item \textbf{MonoT5 (2019)}~\cite{monot5}: a sequence-to-sequence document ranking model that employed T5 architecture.
    \item \textbf{SAILER (2023)}~\cite{SAILER}: a legal structure-aware model that achieves competitive results on the same two datasets.
    \item \textbf{PromptCase (2023)}~\cite{promptcase}: a prompt-based input reformulation method that works on language models for LCR.
    \item \textbf{CaseGNN (2024)}~\cite{casegnn}: a state-of-the-art LCR method using GNN to encode the case text.
\end{itemize}

\subsubsection{Implementation.}
\label{implementation}
The batch size of training are chosen from \{32, 64, 128\}. GAT~\cite{GAT} is the default GNN with number of layers chosen from \{1,2,3\}. The dropout~\cite{dropout} rate is chosen from \{0.1, 0.2, 0.3, 0.4, 0.5\}. Adam~\cite{Adam} is the default optimiser with the learning rate \{1e-4, 1e-5, 1e-6\} and the weight decay \{1e-4, 1e-5, 1e-6\}. For the contrastive training, given a query, the number of positive sample, easy negative sample are both 1 while the number of hard negative samples is chosen from \{1, 5, 10\}. The in-batch samples of other queries are also considered as easy negative samples. For the degree regularisation, $\lambda$ is chosen from \{0,5e-4,1e-3,5e-3\}. CaseGNN~\cite{casegnn} is chosen as the case encoder and SAILER~\cite{SAILER} is chosen as the charge encoder. The two-stage experiment is based on the top 10 BM25 ranking cases as the first stage ranking result. The two-stage experiment results are only conducted for the overall comparison to verify the LCR performance. For all other experiments, only one-stage results are reported. The number of TopK case neighbour node $K$ in Equation~\ref{eq:d-edge} is chosen from \{3, 5, 10, 20\}. The threshold $\delta$ in Equation~\ref{eq:c-edge} is chosen from \{0.85, 0.9, 0.95\}.

\subsection{Overall Performance (RQ1)}
\label{overall}
The overall performance of all baselines and CaseLink is evaluated on COLIEE2022 and COLIEE2023, as shown in Table~\ref{tab:overall}. According to the table, CaseLink achieves state-of-the-art performance by a significant margin compared to all baseline models in both one-stage and two-stage experiments for both COLIEE2022 and COLIEE2023. The improvement margin over the best baseline on COLIEE2022 is relatively smaller than in COLIEE2023. This is because in COLIEE2022, BM25 falls behind with a large margin compared with recent methods. But BM25 is a crucial criteria for the case-case edges, which provides a smaller improvement. While in COLIEE2023, BM25 can achieve a closer performance to other recent baselines, which indicates that BM25 can provide more meaningful edges to boost the performance of CaseLink.

Under the one-stage setting, CaseLink significantly improves LCR performance over baseline models. The strong traditional statistical model BM25 achieves a reasonably strong performance compared with most neural baselines. However, it is not comparable with CaseGNN and CaseLink. The legal corpus pre-trained LEGAL-BERT model is not effective in tackling the comprehensive LCR task compared with other retrieval oriented methods. MonoT5 model gets the worst performance in the overall experiment. One possible reason is that MonoT5 is pre-trained to deal with the text-to-text task while LCR task is different from it. As a BERT-based model, SAILER improves over previous language-based models with its structure aware design. The performance of PromptCase is the best among the language models yet it still falls behind compared with CaseGNN and CaseLink, which indicates that the LCR problem has specific structural and connectivity information in addition to the pure text information. CaseGNN is one of the best performing baseline in LCR, which uses the structural information within the legal case and utilise a GNN to encode the structural information. It outperforms the previous language model-based baselines. As the initialisation of the case node feature for CaseLink, CaseGNN cannot compete with the proposed CaseLink method.

For the two-stage setting, all methods use top10 first-stage results from BM25, followed by a re-ranking using the corresponding baseline models. The two-stage re-ranking experiments are conducted on SAILER, PromptCase, CaseGNN for the comparable one-stage retrieval performance to CaseLink. CaseLink outperforms all the baseline methods in most metrics in two-stage experiments. In COLIEE2022, although CaseLink can outperform other baselines in two-stage ranking, the overall performance is not comparable with one-stage ranking. This is because the one stage ranking of these methods is in a much higher quality than BM25, which limits the ranking candidates obtained from BM25's first stage ranking results. While in COLIEE2023, the two stage ranking can further improve the neural rankers in the ranking-based metrics such as MRR, MAP and NDCG. This is because with a fix selection of correct retrieved cases, CaseLink can rank these cases in a higher position.

\begin{table*}[!t]\centering
\caption{Ablation study. (\%)}\label{tab:ablation}
\resizebox{1\linewidth}{!}{
\begin{tabular}{l|ccccccc|ccccccc}
\toprule
\multirow{2}{*}{Methods} &\multicolumn{7}{|c|}{COLIEE2022} &\multicolumn{7}{c}{COLIEE2023}\\
\cmidrule{2-15}
&P@5 &R@5 &Mi-F1 &Ma-F1 &MRR@5 &MAP &NDCG@5 &P@5 &R@5 &Mi-F1 &Ma-F1 &MRR@5 &MAP &NDCG@5 \\
\midrule\midrule
CaseLink node feat.&35.5$\pm$0.2 &42.1$\pm$0.2 &38.4$\pm$0.3 &42.4$\pm$0.1 &66.8$\pm$0.8 &64.4$\pm$0.9 &69.3$\pm$0.8 &17.7$\pm$0.7 &32.8$\pm$0.7&23.0$\pm$0.5 &23.6$\pm$0.5 &38.9$\pm$1.1 &37.7$\pm$0.8 &42.8$\pm$0.7\\
CaseLink case-case&36.2$\pm$0.3&43.0$\pm$0.3&39.4$\pm$0.3&43.4$\pm$0.3&65.1$\pm$0.4&62.6$\pm$0.3&68.3$\pm$0.3&20.4$\pm$0.1&37.8$\pm$0.1&26.5$\pm$0.1&27.3$\pm$0.1&42.9$\pm$0.5&41.5$\pm$0.4&47.0$\pm$0.5\\
w/o charge-charge&36.1$\pm$0.3&42.9$\pm$0.4&39.2$\pm$0.3&43.2$\pm$0.4&65.9$\pm$0.3&63.8$\pm$0.2&69.4$\pm$0.2&20.8$\pm$0.3&38.7$\pm$0.6&27.1$\pm$0.4&27.9$\pm$0.5&42.2$\pm$0.1&40.6$\pm$0.3&46.2$\pm$0.3\\
w/o residual&20.0$\pm$0.1&23.8$\pm$0.1&21.7$\pm$0.1&23.4$\pm$0.3&36.4$\pm$0.4&35.4$\pm$0.5&40.6$\pm$0.2&20.7$\pm$0.3&38.5$\pm$0.6&27.0$\pm$0.4&27.9$\pm$0.5&44.7$\pm$1.3&43.3$\pm$1.5&48.7$\pm$1.4\\
w/o DegReg&35.9$\pm$0.2&42.7$\pm$0.2&39.0$\pm$0.2&43.1$\pm$0.2&66.0$\pm$0.1&63.8$\pm$0.1&69.5$\pm$0.2&20.5$\pm$0.1&38.1$\pm$0.2&27.1$\pm$0.3&27.9$\pm$0.1&43.6$\pm$0.4&42.1$\pm$0.4&46.7$\pm$0.1\\
\rowcolor{lightgray}
CaseLink &37.0$\pm$0.1&43.9$\pm$0.1&40.1$\pm$0.1&44.2$\pm$0.1&67.3$\pm$0.5&65.0$\pm$0.2&70.3$\pm$0.1&20.9$\pm$0.3&38.4$\pm$0.6&27.1$\pm$0.3&28.2$\pm$0.3&45.8$\pm$0.5&44.3$\pm$0.7&49.8$\pm$0.4\\
\bottomrule
\end{tabular}}
\end{table*}

\begin{table*}[!t]\centering
\caption{Effectiveness of Case Feature Initialisation. (\%)}\label{tab:effect_init}
\resizebox{1\linewidth}{!}{
\begin{tabular}{l|ccccccc|ccccccc}
\toprule
\multirow{2}{*}{Methods} &\multicolumn{7}{|c|}{COLIEE2022} &\multicolumn{7}{c}{COLIEE2023}\\
\cmidrule{2-15}
&P@5 &R@5 &Mi-F1 &Ma-F1 &MRR@5 &MAP &NDCG@5 &P@5 &R@5 &Mi-F1 &Ma-F1 &MRR@5 &MAP &NDCG@5 \\
\midrule\midrule
PromptCase &17.1 &20.3 &18.5 &20.5 &35.1 &33.9 &38.7 &16.0 &29.7 &20.8 &21.5 &32.7 &32.0 &36.2\\
\rowcolor{lightgray}
+CaseLink&18.6$\pm$0.2&22.1$\pm$0.3&20.2$\pm$0.3&22.0$\pm$0.3&39.0$\pm$0.1&37.3$\pm$0.1&41.7$\pm$0.1&19.5$\pm$0.2&36.1$\pm$0.4&25.3$\pm$0.3&26.1$\pm$0.3&42.4$\pm$0.5&41.2$\pm$0.5&46.4$\pm$0.5\\
\midrule
\midrule
CaseGNN &35.5$\pm$0.2 &42.1$\pm$0.2 &38.4$\pm$0.3 &42.4$\pm$0.1 &66.8$\pm$0.8 &64.4$\pm$0.9 &69.3$\pm$0.8 &17.7$\pm$0.7 &32.8$\pm$0.7&23.0$\pm$0.5 &23.6$\pm$0.5 &38.9$\pm$1.1 &37.7$\pm$0.8 &42.8$\pm$0.7\\
\rowcolor{lightgray}
+CaseLink&37.0$\pm$0.1&43.9$\pm$0.1&40.1$\pm$0.1&44.2$\pm$0.1&67.3$\pm$0.5&65.0$\pm$0.2&70.3$\pm$0.1&20.9$\pm$0.3&38.4$\pm$0.6&27.1$\pm$0.3&28.2$\pm$0.3&45.8$\pm$0.5&44.3$\pm$0.7&49.8$\pm$0.4\\
\bottomrule
\end{tabular}}
\end{table*}

\begin{table*}[!t]\centering
\caption{Effectiveness of Case-Case Edges. (\%)}\label{tab:effect_caseedge}
\resizebox{1\linewidth}{!}{
\begin{tabular}{l|ccccccc|ccccccc}
\toprule
\multirow{2}{*}{Methods} &\multicolumn{7}{|c|}{COLIEE2022} &\multicolumn{7}{c}{COLIEE2023}\\
\cmidrule{2-15}
&P@5 &R@5 &Mi-F1 &Ma-F1 &MRR@5 &MAP &NDCG@5 &P@5 &R@5 &Mi-F1 &Ma-F1 &MRR@5 &MAP &NDCG@5 \\
\midrule\midrule
CaseGNN-Edge &35.8$\pm$0.1&42.5$\pm$0.1&38.8$\pm$0.1&43.1$\pm$0.1&64.4$\pm$0.1&62.1$\pm$0.1&67.9$\pm$0.1&17.3$\pm$0.1&32.2$\pm$0.1&22.5$\pm$0.1&23.3$\pm$0.1&36.6$\pm$0.5&35.5$\pm$0.3&40.7$\pm$0.4\\
\rowcolor{lightgray}
BM25-Edge&37.0$\pm$0.1&43.9$\pm$0.1&40.1$\pm$0.1&44.2$\pm$0.1&67.3$\pm$0.5&65.0$\pm$0.2&70.3$\pm$0.1&20.9$\pm$0.3&38.4$\pm$0.6&27.1$\pm$0.3&28.2$\pm$0.3&45.8$\pm$0.5&44.3$\pm$0.7&49.8$\pm$0.4\\
\bottomrule
\end{tabular}}
\end{table*}

\begin{table*}[!t]\centering
\caption{Effectiveness of GNNs. (\%)}\label{tab:effect_gnn}
\resizebox{1\linewidth}{!}{
\begin{tabular}{l|ccccccc|ccccccc}
\toprule
\multirow{2}{*}{Methods} &\multicolumn{7}{|c|}{COLIEE2022} &\multicolumn{7}{c}{COLIEE2023}\\
\cmidrule{2-15}
&P@5 &R@5 &Mi-F1 &Ma-F1 &MRR@5 &MAP &NDCG@5 &P@5 &R@5 &Mi-F1 &Ma-F1 &MRR@5 &MAP &NDCG@5 \\
\midrule\midrule
GCN &34.5$\pm$0.1&41.0$\pm$0.1&37.5$\pm$0.1&41.4$\pm$0.1&66.0$\pm$0.1&63.5$\pm$0.2&69.1$\pm$0.2&18.1$\pm$0.1&33.6$\pm$0.1&23.5$\pm$0.1&24.4$\pm$0.1&39.5$\pm$0.1&38.2$\pm$0.1&43.7$\pm$0.1\\
GraphSAGE &34.6$\pm$0.1&41.1$\pm$0.2&37.6$\pm$0.1&41.7$\pm$0.1&64.4$\pm$0.3&62.3$\pm$0.2&67.8$\pm$0.4&18.7$\pm$0.4&34.8$\pm$0.8&24.3$\pm$0.5&25.3$\pm$0.6&39.6$\pm$0.7&38.2$\pm$0.6&43.6$\pm$0.9\\
GAT &37.0$\pm$0.1&43.9$\pm$0.1&40.1$\pm$0.1&44.2$\pm$0.1&67.3$\pm$0.5&65.0$\pm$0.2&70.3$\pm$0.1&20.9$\pm$0.3&38.4$\pm$0.6&27.1$\pm$0.3&28.2$\pm$0.3&45.8$\pm$0.5&44.3$\pm$0.7&49.8$\pm$0.4\\
\bottomrule
\end{tabular}}
\end{table*}

\subsection{Ablation Study (RQ2)}
The ablation study is conducted to validate the effectiveness of different modules in CaseLink: (1) CaseLink node feat., which is the initialised node features of CaseLink without using any connection relationship; (2) CaseLink case-case, which is equivalent to only use the case-case edge in addition to the node features; (3) w/o charge-charge, just removing the charge-charge cases from CaseLink; (4) w/o residual, not using the residual connection in GNN; (5) w/o DegReg, optimising the CaseLink without the degree regularisation of candidates. All experiments are conducted on both datasets and evaluated on all metrics for one-stage retrieval, as in Table~\ref{tab:ablation}.

As shown in Table~\ref{tab:ablation}, the complete CaseLink can outperform all other variants. The CaseLink node feat.\ uses node features initialised from CaseGNN as the baseline. For the CaseLink case-case variant, there is a significant improved performance over in all the metrics for COLIEE2023 and non-ranking metrics for COLIEE2022, which indicates the effectiveness of case connectivity relationship. For the variant w/o charge-charge, the performance is lower than CaseLink in most metrics except the non-ranking ones in COLIEE2023, implying the charge-charge edges can provide further information during graph learning between charges. For w/o residual, the performance decreases for both COLIEE2022 and COLIEE2023, which maybe because without the high quality node initialisation, the LCR would be largely affected by the BM25 case-case edges. For w/o DegReg, the performance is worse than CaseLink, which proves that using degree regularisation can effectively reduce useless connection of candidates in the case pool.

\subsection{Effectiveness of Graph Learning (RQ3)}
\subsubsection{Different Case Feature Initialisation Strategies}
To explore the impact of different initialisation strategies to CaseLink, the experiment uses two different types of node initialisation from PromptCase and CaseGNN. The one-stage result are shown in Table~\ref{tab:effect_init} and evaluated on all metrics for both datasets. According to the results, it is obvious that CaseLink can improves the performance on both PromptCase and CaseGNN, which indicates the effectiveness of graph learning based on case connectivity relationships and degree regularisation for LCR performance. 

\subsubsection{Different Case-Case Edge Selections}
To evaluate the impact of different case-case edge selections, this experiment of choose to use CaseGNN embedding to generate the case-case edges and compare the performance with the original BM25-based edges. As shown in Table~\ref{tab:effect_caseedge}, the performance of BM25-Edge is better than CaseGNN-Edge, which maybe because using BM25 similarity score to select the edges can provide a different type of information from statistical-based similarity to the CaseLink, while CaseGNN can only provide the similarity information from the case embedding, which is already included as the node input features by CaseLink.

\subsubsection{Different GNN Layers}
In this experiment, different choices of GNN layers are evaluated by replacing the default GAT with GCN~\cite{GCN} and GraphSAGE~\cite{GraphSAGE}. All experiments are conducted on both datasets and evaluated on all metrics in one-stage setting as shown in Table~\ref{tab:effect_gnn}. According to the results, GAT achieves the highest performance among other two widely used GNN models, GCN and GraphSAGE. This phenomenon is aligned with the performance gap on other general graph learning tasks for the different graph learning ability among GAT, GCN and GraphSAGE.

\subsection{Parameter Sensitivity (RQ4)}
\label{sec:parameter}
In this experiment, the DegReg coefficient $\lambda$ in Equation~(\ref{eq:loss-overall}), the number $K$ of TopK in Equation~(\ref{eq:d-edge}) and the number $k$ of GNN layers are studied for the parameter sensitivity of CaseLink.

\begin{figure}[!t]
\centering
    \subfigure{
    \includegraphics[width=0.47\linewidth]{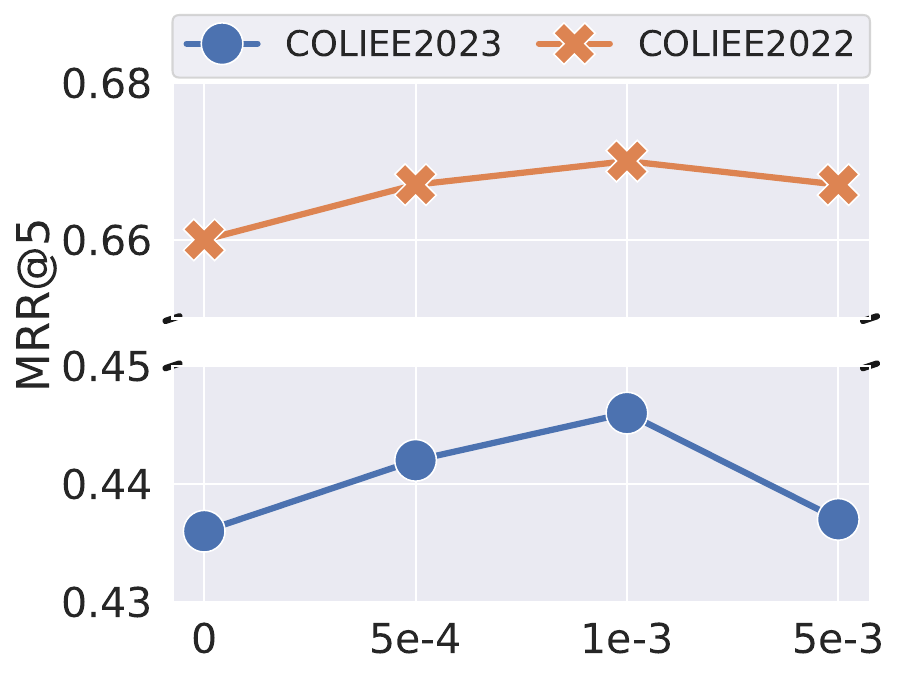}
    \label{fig:lamb-mrr}
    }
    \subfigure{
    \includegraphics[width=0.47\linewidth]{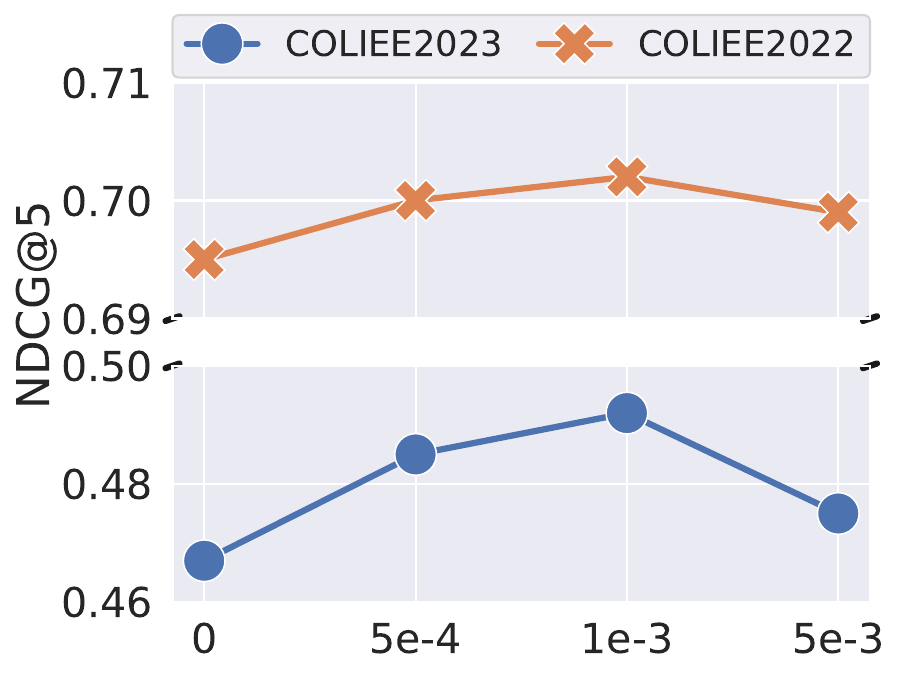}
    \label{fig:lamb-ndcg}
    }
    \vspace{-0.3cm}
\caption{Parameter sensitivity of $\lambda$ in Equation~(\ref{eq:loss-overall}).}
\label{fig:lamb}
\end{figure}

\subsubsection{DegReg Coefficient $\lambda$.} As shown in Figure~\ref{fig:lamb}, $\lambda$ is chosen from \{0, 5e-4, 1e-3, 5e-3\}. $\lambda$ set to 1e-3 achieves the best performance on both datasets. When $\lambda$ is too small, the training loss of DegRes is decreased and the training signal for candidates is not enough, which results in a worse performance. On the contrary, when $\lambda$ is too large, the overall loss of CaseLink is paying too much attention to the DegReg loss while ignoring the contrastive loss for training query, which will lead to insufficient training of the LCR task.

\subsubsection{TopK BM25 Neighbours in Case-Case Edge.} 
\begin{figure}[!t]
\vspace{-0.3cm}
\centering
    \subfigure{
    \includegraphics[width=0.47\linewidth]{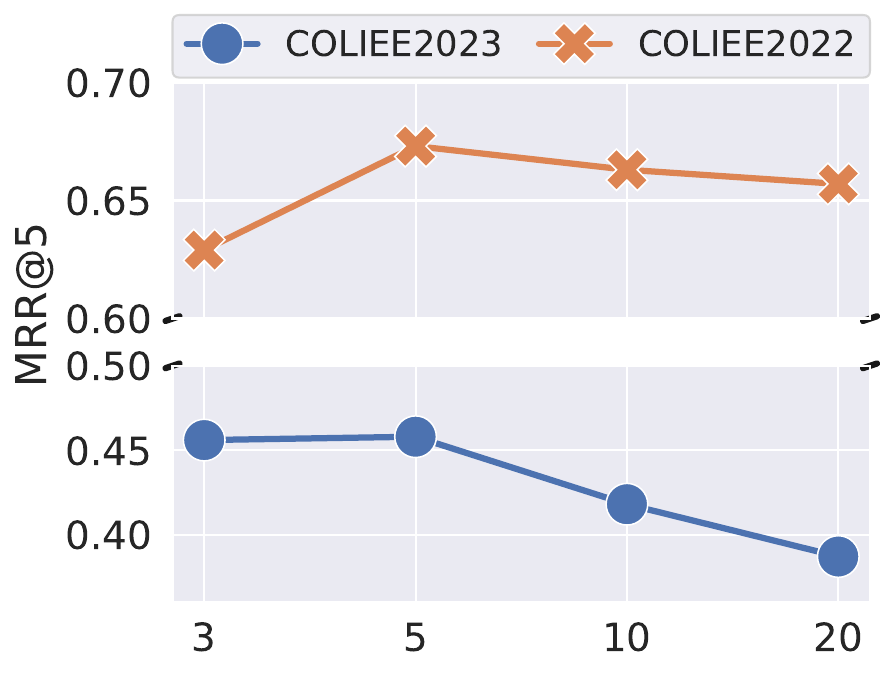}
    \label{fig:topkneibor_mrr}
    }
    \subfigure{
    \includegraphics[width=0.47\linewidth]{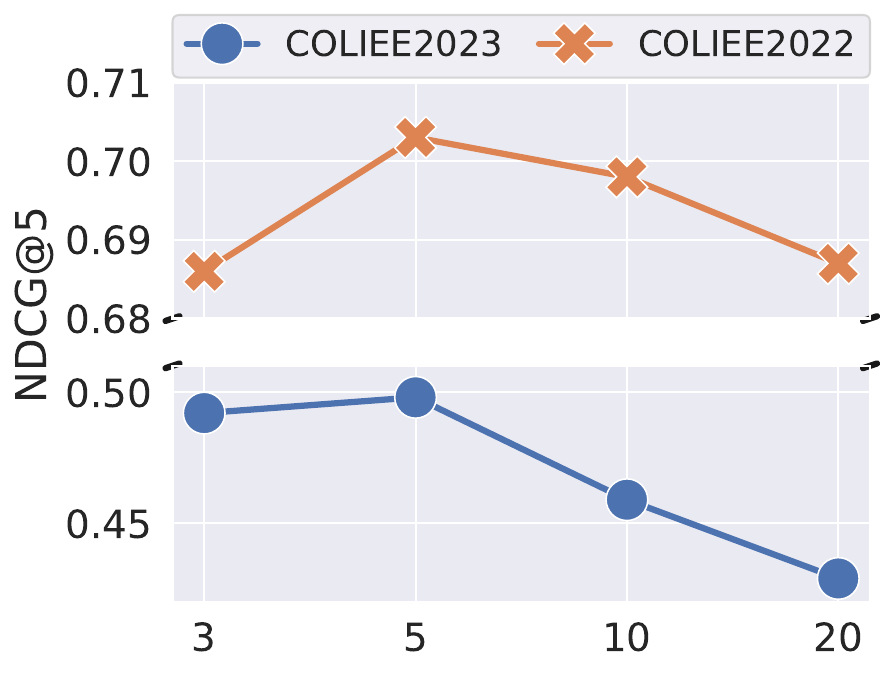}
    \label{fig:topkneibor_ndcg}
    }
    \vspace{-0.3cm}
\caption{Parameter sensitivity of TopK BM25 neighbours in case-case edge construction from Equation~(\ref{eq:d-edge}).}
\vspace{-0.5cm}
\label{fig:topkneibor}
\end{figure}

According to Figure~\ref{fig:topkneibor}, the choice of number of TopK neighbour cases are from \{3, 5, 10, 20\}. The TopK neighbour cases are sampled from the ranking list of BM25 similarity score. Different TopK neighbour cases numbers in Equation~(\ref{eq:d-edge}) have different impacts to the generated CGC.
As shown in Figure~\ref{fig:topkneibor}, when the number of neighbours is increased, the performance is getting worse, which maybe because the average number of ground truth label in both datasets are closer to 5 as shown in Table~\ref{tab:dataset}. During training, nodes in GCG with a similar neighbour number to ground truth label number can make the graph learning for LCR easier and quicker. What's more, using similar number of neighbours for constructing GCG can get more case connectivity relationships for the similar node numbers and edges numbers to the ground truth label graph. While the number of neighbours is too small such as 3, the performance also becomes worse because of the lacking enough nodes information and case connectivity information within the graph.

\subsubsection{GNN layer numbers}
\label{sec:layer}
\begin{figure}[!t]
\centering
    \subfigure{
    \includegraphics[width=0.47\linewidth]{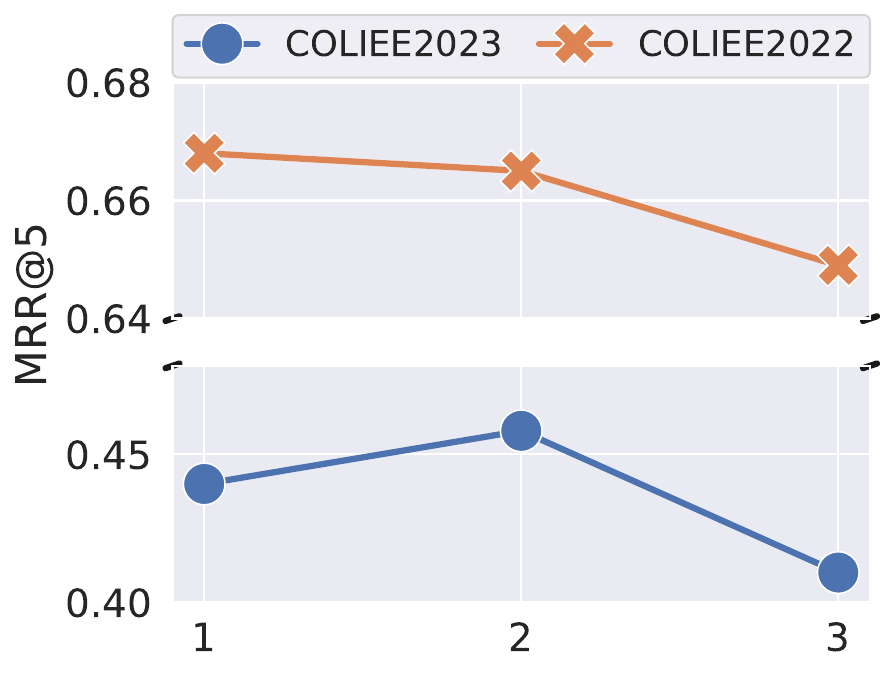}
    \label{fig:layer-mrr}
    }
    \subfigure{
    \includegraphics[width=0.47\linewidth]{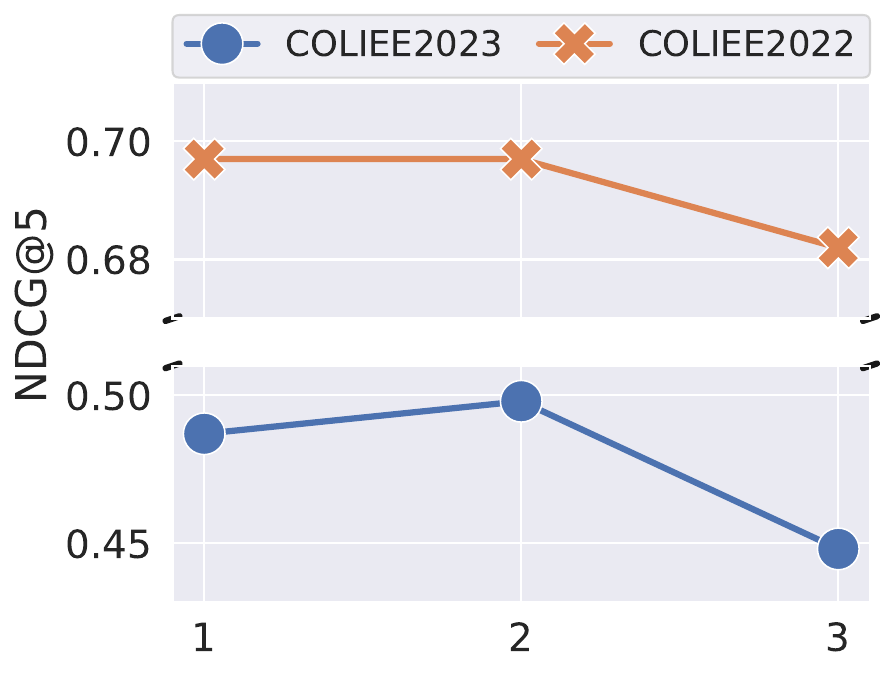}
    \label{fig:layer-ndcg}
    }
    \vspace{-0.3cm}
\caption{Parameter sensitivity of GNN layer number.}
\label{fig:layer}
\end{figure}
In Figure~\ref{fig:layer}, the number of GNN layers are chosen from \{1, 2, 3\}. The 2-layer GNN model is the best on COLIEE2023 while on COLIEE2022, the 1-layer and 2-layer models are better than 3-layer. This phenomenon shows that different datasets have different sensitivities on GNN layer number.

\begin{figure}[!t]
\centering
\includegraphics[width=0.8\linewidth]{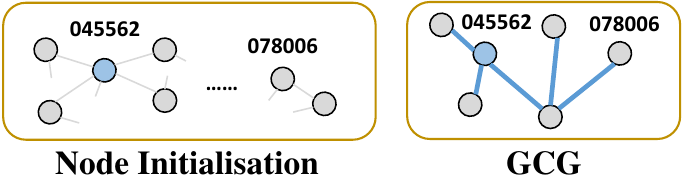}
\vspace{-0.2cm}
\caption{CaseLink using high-order case-case edge to correctly retrieve candidate `078006' for query `045562'.}
\vspace{-0.3cm}
\label{fig:casestudy_no_charge}
\end{figure}

\begin{figure}[!t]
\centering
\includegraphics[width=0.8\linewidth]{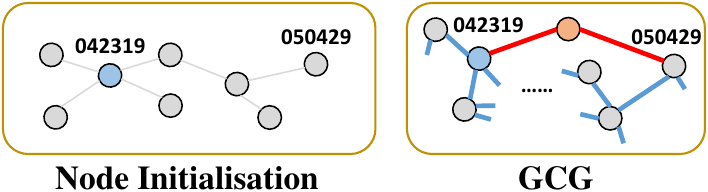}
\vspace{-0.2cm}
\caption{CaseLink using high-order case-charge edge to correctly retrieve candidate `050429' for query `042319'.}
\vspace{-0.5cm}
\label{fig:casestudy_charge}
\end{figure}

\subsection{Case Study}
This case study provides two examples of how the graph learning helps with CaseLink to conduct effective LCR in Figure~\ref{fig:casestudy_no_charge} and~\ref{fig:casestudy_charge}. The blue node is the query case and grey nodes are candidate cases. Light grey edges are only used to indicate similarity of input node features, which is not used in CaseLink. Blue edges and red edges are from GCG, which are used in CaseLink.

In Figure~\ref{fig:casestudy_no_charge}, the query case `045562' and the ground truth candidate case `078006' are faraway in the node feature space. The case-case edges based on BM25 scores in GCG bring these two nodes into a two-hop neighbourhood. CaseLink successfully utilise this intrinsic connectivity to make the correct prediction.

In Figure~\ref{fig:casestudy_charge}, the query case `042319' and the ground truth candidate case `050429' are in three-hop neighbourhood, not close enough to make the correct prediction directly using node features. This is also reflected by that there is no close connection with case-case edges. However, these two cases are both connected to a charge node by case-charge edges, which makes these two cases within two-hop neighbour in GCG. CaseLink successfully utilise this intrinsic connectivity to make the correct prediction.

\section{Conclusion}
This paper focuses on leveraging the case intrinsic connectivity in legal case retrieval. To achieve this goal, a CaseLink model is proposed, consisting of a Global Case Graph construction module to provide connections between cases and charges, and a degree regularisation to provide training signals for candidate cases. Extensive experiments conducted on two benchmark datasets verify the state-of-the-art performance and the effectiveness of CaseLink.

\section{Acknowledgements} This work is supported by Australian Research Council
CE200100025, FT210100624, DP230101196, DP230101753, DP240101108.

\bibliographystyle{ACM-Reference-Format}
\bibliography{sample-base.bib}

\end{document}